\documentclass[aps,prl,superscriptaddress, twocolumn,a4paper,showpacs]{revtex4-1}

\usepackage{graphicx}
\usepackage{setspace}
\usepackage[dvips,usenames]{color}

\newcommand{\eg}{{\it e.g.}}
\newcommand{\ie}{{\it i.e.}}
\newcommand{\bra}[1]{{\langle #1 |}}
\newcommand{\ket}[1]{{| #1 \rangle}}

\begin{document}

\title{Symmetry breaking between statistically equivalent, independent
  channels in a few-channel chaotic scattering}

\author{C. Mej\'{\i}a-Monasterio}
\email{carlos.mejia@upm.es}
\affiliation{Laboratory of Physical Properties,
Technical University of Madrid, Av. Complutense s/n, 28040 Madrid, Spain}
\author{G. Oshanin}
\email{oshanin@lptmc.jussieu.fr}
\affiliation{Laboratoire de Physique Th{\'e}orique de la Mati{\`e}re
Condens{\'e}e (UMR CNRS 7600), Universit{\'e} Pierre et Marie Curie (Paris 6) -
4 Place Jussieu, 75252 Paris, France}

\author{G. Schehr}
\email{gregory.schehr@th.u-psud.fr}
\affiliation{Laboratoire de Physique Th\'eorique, Universit\'e de Paris-Sud, France}

\date{\today}


\begin{abstract}
  We study the distribution function $P(\omega)$ of the random
  variable $\omega = \tau_1/(\tau_1 + \ldots + \tau_N)$, where
  $\tau_k$'s are the partial Wigner delay times for chaotic scattering
  in a disordered system with $N$ independent, statistically
  equivalent channels. In this case, $\tau_k$'s are i.i.d. random
  variables with a distribution $\Psi(\tau)$ characterized by a "fat"
  power-law intermediate tail $\sim 1/\tau^{1 + \mu}$, truncated by an
  exponential (or a log-normal) function of $\tau$. For $N = 2$ and $N
  = 3$, we observe a surprisingly rich behavior of $P(\omega)$
  revealing a breakdown of the symmetry between identical independent
  channels. For $N=2$, numerical simulations of the quasi
  one-dimensional Anderson model confirm our findings.
\end{abstract}

%\pacs{02.50.-r; 03.65.Nk; 42.25.Dd; 73.23.-b}

\maketitle

Scattering by chaotic or disordered systems is encountered in various
situations ranging from nuclear, atomic or molecular physics to
mesoscopic devices \cite{yan1}.  The key property characterizing the
scattering process is the unitary $S$ matrix relating the amplitudes
of waves incoming to the system and those of scattered waves. Since
the underlying dynamics is chaotic, the properties of the $S$ matrix
behave in an irregular way when the parameters of the incoming waves
or of the medium are modified. Hence, an adequate description of the
scattering process requires the knowledge of the $S$ matrix
distribution.

Time-dependent aspects of the scattering process are well captured by
the Wigner delay time (WDT) $\tau$ \cite{wigner}, defined through the
derivative of the $S$ matrix with respect to energy $E$. Physically,
$\tau$ is an excess time spent in the interaction region by a wave
packet with energy peaked at $E$, as compared to a free wave packet
propagation.

For systems coupled to the outside world via $N$ open channels, $\tau
= \sum_{k=1}^N \tau_k$, where $\tau_k = \partial \theta_k/\partial E$
are the \textit{partial} delay times and $\theta_k$ are the phase
shifts of the $S$ matrix. One shows that $\tau_k$'s are the diagonal
elements of the Wigner-Smith time delay matrix (WSM), taken in the
eigenbasis of the scattering matrix \cite{corr}.  Note that it is also
customary to define the eigenvalues of the WSM as the \textit{proper}
delay times $\tau_k'$ (see, \eg, Ref.~\cite{yan1,beenaker}).  Beyond
the $1$-channel case, $\tau_k$ and $\tau_k'$ differ, although their
sums over all scattering channels are equal to each other.

Likewise the $S$-matrix, the WDT is a random
variable, whose distribution $\Psi(\tau)$ has a
generic form
\cite{comtet2,bolton,ossipov,steinbach,geisel,fyod,fJLett,sym,2,3,corr}:
 \begin{equation}
\label{distribution}
\Psi(\tau) = \frac{a^{\mu}}{\Gamma(\mu)} \exp\left(- \frac{a}{\tau}\right) \, \frac{1}{\tau^{1 + \mu}} \, ,
\end{equation}
where $a$ is a characteristic parameter, $\Gamma(\mu)$ is the gamma
function and $\mu$ is a model-dependent exponent: one encounters
situations with $0 < \mu < 1$, $\mu = 1$ and $\mu > 1$.

For 1D single-channel systems with weak disorder $\mu = 1$
\cite{comtet2,bolton,ossipov}, which holds also for quasi-1D
disordered systems of length $L \gg \lambda$, where $\lambda$ is the
localization length \cite{fJLett}.  One can demonstrate the validity of this result for a
single-channel scattering in any dimension in the regime of strong
localization \cite{3}.
In 1D quasi-periodic systems with a single open channel and fractal
dimension $D_0^E$ ($\leq 0.5$) of the spectrum one has
 $\mu = 1 - D_0^E < 1$ \cite{steinbach}, and
$\mu = 1/2$ holds for the 2D generalization of a kicked
rotor model \cite{fyod,geisel}, as well as for generic weakly open chaotic
systems in a parametrically large range of delay times
\cite{yan1,fyod}.
Lastly, $\mu = 1 + N \beta/2 > 1$, where $\beta$ is
the Dyson symmetry index, was obtained for ballistic scattering from a
cavity \cite{sym,2,3}.

It is however clear that Eq. (\ref{distribution}) defines a limiting
form, valid either for $L \to \infty$ or for weakly open systems. In
reality, the power-law tail is truncated, such that all moments of
$\Psi(\tau)$ exist.  Two model-dependent cut-offs seem to be
physically plausible (although not exact) \cite{comtet2}:
\begin{equation}
\label{atruncated_dist}
\Psi_1(\tau) = \frac{1}{2}  \frac{\left(a b\right)^{\mu/2}}{K_{\mu}(2 \sqrt{a/b})} \exp\left(- \frac{a}{\tau}\right) \, \frac{1}{\tau^{1 + \mu}} \exp\left( - \frac{\tau}{b}\right) \;,
\end{equation}
$K_\mu(x)$ being the modified Bessel function, and a log-normally truncated (LNT) form with
$\exp(-\ln^2(\tau)/c)$ in place of  $\exp(-\tau/b)$, where $b$ and $c$ are either
$\sim L$ \cite{yan1,comtet2}, or to the opening
degree for weakly open systems \cite{yan1,fyod}.

In this paper we are concerned with a somewhat unusual statistics of
partial delay times for scattering in systems with $N$
\textit{equivalent} channels. We focus here on
\begin{equation}
\label{omega}
\omega = \frac{\tau_1}{\tau_1 + \tau_2 + ... + \tau_N},
\end{equation}
a random variable which probes the \textit{contribution} of one of the
channels to the WDT and hence, the symmetry
between different channels.  To highlight the effect of the
intermediate power-law tail of $\Psi(\tau)$, we suppose that the
channels are independent of each other such that the partial delay
times $\tau_k$'s are i.i.d. random variables with a common
distribution in Eqs. (\ref{distribution}) or (\ref{atruncated_dist}) (or a LNT form).  This
situation can be realized experimentally, \eg, for scattering in a
bunch of disordered fibers.  Such a simplified model
with $\mu=1$ is also appropriate for a multichannel scattering from a
piece of strongly disordered media when the distance between the
scattering channels locations exceeds $\lambda$.  The role of
correlations will be briefly discussed at the end of this paper.

We show here, on example of $2$- and $3$-channel systems, that
\textit{intermediate} power-law tails entail a surprisingly rich
behavior of the distribution
 \begin{eqnarray}
\label{dist}
P(\omega) = \left<\delta\left(\omega - {\tau_1}/{(\tau_1 + \tau_2 + ... + \tau_N)}\right)\right>,
\end{eqnarray}
where $\langle \dots \rangle$ denotes an average over the
distributions of $\tau_k$'s.  We realize that $\omega$ exhibits
significant sample-to-sample fluctuations and, in general, the
symmetry between identical independent channels is broken, despite the
fact that all the moments of $\Psi(\tau)$ are well defined. A similar
result was found for related mathematical objects in
Refs.~\cite{redner,we_jstat}.  We address the reader to
Ref.~\cite{we_jstat} for the details on the derivation of $P(\omega)$.

For $N = 2$ and $\Psi(\tau)$ as in
Eq. (\ref{distribution}), we get:
\begin{equation} \label{lim}
P(\omega) \equiv B \, \omega^{\mu - 1} \left(1 - \omega\right)^{\mu - 1}.
\end{equation}
with $B = {\Gamma(2 \mu)}/{\Gamma^2(\mu)}$.
\begin{figure}[b]
  \centerline{\includegraphics*[width=\linewidth]{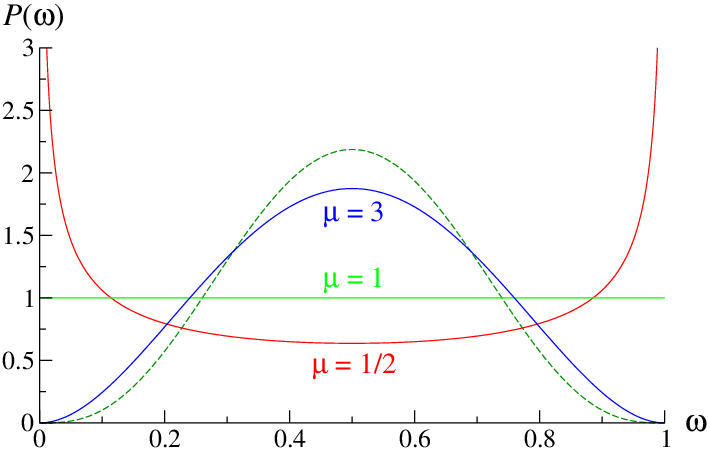}}
\caption{(Color online) $N = 2$. $P(\omega)$ in Eq. (\ref{lim}) for different values of
$\mu$. The dashed line depicts $P(\omega)$, Eq.~(\ref{cor}), with $\beta = 2$ $(\mu = 4)$.
}
  \label{sketch}
\end{figure}
A striking feature of the beta-distribution in Eq. (\ref{lim}) is that
its very shape depends on whether $0 < \mu < 1$, $\mu = 1$
or $\mu > 1$ (see Fig.~\ref{sketch}).  For $0 < \mu < 1$, $P(\omega)$
is bimodal with a $U$-like shape, and most probable values
being $0$ and $1$.  In this case, the
symmetry between two identical independent channels is broken and
either of the two channels provides a dominant contribution to
the WDT. Strikingly, $\langle \omega \rangle =
1/2$ corresponds here to the \textit{least} probable value of
$\omega$.  For $\mu = 1$, $P(\omega) \equiv 1$, and either of the
channels may provide \textit{any contribution} to the overall delay
time with \textit{equal} probability.  Finally, for $\mu > 1$,
$P(\omega)$ is unimodal, which signifies
that both channels contribute proportionally.

\begin{figure}[ht]
  \centerline{\includegraphics[width=\linewidth]{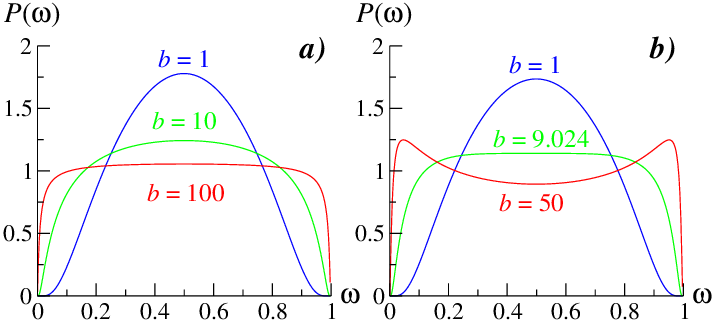}}
  \caption{(Color online) $P(\omega)$ for $N = 2$ and $\Psi(\tau)$ in Eq. (\ref{atruncated_dist}) with $a = 1$. {\bf a)}: for $\mu =
    1$ and different $b$. {\bf b)}: for $\mu = 1/2$ and
    different $b$.  }
  \label{sketch2}
\end{figure}

For $N = 2$ and a truncated $\Psi(\tau)$ as in
Eq. (\ref{atruncated_dist}) we get
\begin{equation}
\label{9}
 P(\omega) = B [\omega (1 - \omega)]^{-1} {K_{2 \mu}\left(2 \sqrt{a/b \, \omega (1 - \omega)}\right)} \,.
\end{equation}
where $B = [{2 K_{\mu}^2(2 \sqrt{a/b})}]^{-1}$. Note that $P(\omega)$
vanishes at the edges and is symmetric around $\omega = 1/2$. The
behavior of $P(\omega)$ can be analyzed by expanding the expression in
Eq. (\ref{9}) in Taylor series at $\omega = 1/2$ \cite{we_jstat}.
%\begin{eqnarray}
%\label{exp}
%&&P(\omega) \sim 1 + g \, \left(\omega - {1}/{2}\right)^2 + {\mathcal O}\left((\omega - {1}/{2})^4\right) \, \label{exp} \\
%&& g \propto 1 - \mu - 2 \sqrt{\frac{a}{b}} \frac{K_{2 \mu - 1}(4 \sqrt{a/ b})}{K_{2 \mu}(4 \sqrt{a/ b})} \, . \label{g}
%\end{eqnarray}
For $\mu > 1$, $P(\omega)$ is a bell-shaped function with a
\textit{maximum} at $\omega = 1/2$. For $\mu = 1$, for which we
previously found a uniform distribution, the latter (apart from an
exponential cut-off at the edges) is approached in the limit $b/a \gg
1$ [see Fig.\ref{sketch2} a)].
For $0 < \mu < 1$ the situation is more complicated: there exists a
critical value $b_c$ separating two different regimes.
%\begin{equation}
%1 - \mu = 2 \sqrt{\frac{a}{b_c}} \frac{K_{2 \mu - 1}(4 \sqrt{a/b_c})}{K_{2 \mu}(4 \sqrt{a/ b_c})}.
%\end{equation}
For $b < b_c$, $P(\omega)$ is a bell-shaped function with a maximum at
$\omega = 1/2$. For $b = b_c$, $P(\omega) \approx 1$ except for narrow
regions at the edges, where it vanishes exponentially. Finally, for $b
> b_c$ $P(\omega)$ has an $M$-like shape, with maxima close to
$\omega = 0$ and $\omega = 1$, $\omega = 1/2$ being the least probable
value.  Hence, an exponential truncation of $\Psi(\tau)$ does not
restore the symmetry between different channels, which holds only for
systems whose size is less than some critical length set by $b_c$.
Note that for a LNT form
the overall behavior
of $P(\omega)$ is very similar and also exhibits a transition at some value of the parameter $c$.

\begin{figure}[ht]
  \centerline{\includegraphics[width=\linewidth]{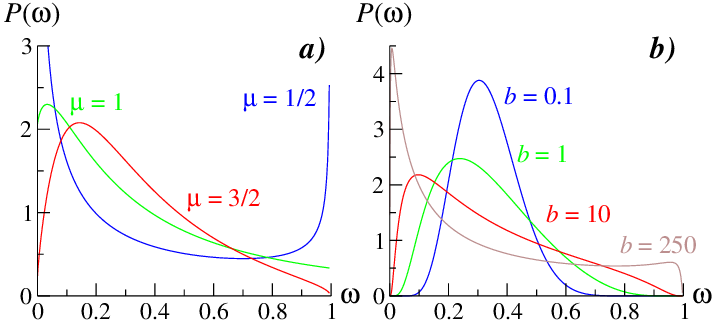}}
  \caption{(Color online) $N = 3$. {\bf a)}: $P(\omega)$ in Eq. (\ref{N3}) for
    different values of $\mu$. {\bf b)}: $P(\omega)$ in Eq. (\ref{k})
    for $\mu=1/2$ and for different values of $b$ ($a$ is set equal to $1$).  }
  \label{sketch5}
\end{figure}

Further on, for $N = 3$ and $\Psi(\tau)$
as in~(\ref{distribution}),
\begin{equation}
\label{N3}
P(\omega) = C   \frac{\left(1 - \omega\right)^{\mu - 1}}{\omega^{1 + \mu}}\,_2F_1\left(2 \mu,
3 \mu; 2 \mu + \frac{1}{2}; \frac{\omega-1}{4 \omega}\right)
\end{equation}
where $C = \frac{\sqrt{\pi}}{2^{4 \mu - 1}} \; \frac{\Gamma(2 \mu)
  \Gamma(3 \mu)}{\Gamma^3(\mu) \Gamma(2 \mu + 1/2)} $ and
$_2F_1$ is a hypergeometric series. One finds from Eq. (\ref{N3}) that
$P(\omega) \sim (1 - \omega)^{\mu - 1}$ when $\omega \to 1$ and
$P(\omega) \sim \omega^{\mu - 1}$ when $\omega \to 0$, which agrees
with the result in Eq. (\ref{lim}). On the other hand, the amplitudes
in these asymptotic forms appear to be very different such that
$P(\omega)$ is skewed to the left [see Fig.~\ref{sketch5} a)].
Therefore, for $N = 3$, $P(\omega)$ diverges at both edges and has a
$U$-shaped form for $0 < \mu < 1$ (see Fig.~\ref{sketch5} ~a), which signifies that
the symmetry between the channels is broken.  For $\mu \geq 1$ the
distribution is unimodal. Remarkably, for $\mu \geq 1$ the maximum $\omega_m$ of $P(\omega)$
is not at $\omega = 1/3$: for $\mu =3$, one has $\omega_m = 0.2719$,
for $\mu = 10$ the maximum is at $\omega_m = 0.3102$ and etc;
actually, $\omega_m \to 1/3$ only when $\mu \to \infty$. This means
that even for $\mu > 1$, $\omega$ exhibits sample to sample
fluctuations and the average value $\langle \omega \rangle \equiv 1/3$
does not have any significance.

For $N = 3$ and a truncated
$\Psi(\tau)$ as in Eq. (\ref{atruncated_dist}) we get
\begin{eqnarray}
\label{k}
P(\omega) &=& \frac{(b/4 a)^{3 \mu/2}}{2 K^3_{\mu}\left(2 \sqrt{a/b}\right)} \frac{\omega^{\mu - 1}}{(1 - \omega)^{\mu + 1}} \int^{\infty}_0 dx \, x^{\mu} \, J_{\mu - 1}\left(x\right) \nonumber\\ &\times& \left(x^2 + \frac{4 a}{b \omega}\right)^{\mu} \, K^2_{\mu}\left(\sqrt{\frac{\omega}{1 - \omega}\left(x^2 + \frac{4 a}{b \omega}\right)} \right),
\end{eqnarray}
where $J_\mu(x)$ is the Bessel function. We observe that $P(\omega)$
in Eq. (\ref{k}) is always a bell-shaped function for $\mu \geq
1$. The most probable value $\omega_m$ is, however, always
substantially less than $1/3$, approaching this value only when $\mu
\to \infty$ or $b \to 0$. The case $0 < \mu < 1$ is
different:
%Focusing on $\mu = 1/2$, for which Eq. (\ref{k}) simplifies,
%
%\begin{equation}
%\label{3exp}
%P(\omega) = \frac{4}{\pi} \sqrt{\frac{a}{b}} \frac{e^{6 \sqrt{a/ b}}}{\omega (1 - \omega)}
%\frac{K_1\left(2 \sqrt{\frac{a}{b} \frac{1 + 3 \omega}{\omega (1- \omega)}}\right)}{\sqrt{1 + 3 \omega}},
%\end{equation}
%we discuss a sequence of different regimes which may be
%observed when $b/a$ is gradually varied [see Fig.~\ref{sketch5} b)].
For $b/a \ll 1$, $P(\omega)$ is peaked at $\omega_m \approx 1/3$. For
larger $b/a$, $\omega_m$ moves towards the origin and $P(\omega_m)$
decreases. For yet larger $b/a$, $\omega_m$ keeps moving towards the
origin but now $P(\omega_m)$ passes through a minimum and then starts
to grow. At some special value of $b/a$ ($b/a \approx 140$ for
$\mu=1/2$) a second extremum emerges (at $\omega \approx 0.84$ for
$\mu =1/2$) which then splits into a minimum and a maximum and
$P(\omega)$ becomes bimodal. For still larger $b/a$, the minimum moves
towards $\omega = 1/2$, while the second maximum moves to $\omega =
1$.
%\begin{figure}[ht]
 % \centerline{\includegraphics*[width=0.35\textwidth]{Exponential_truncation_mu=05_location_extrema.eps}}
%\caption{$N = 3$. The loci of the extrema of $P(\omega)$ in Eq. (\ref{3exp}). Blue (green) circles define the position of the first (second) maximum. Red circles define the position of the minimum of $P(\omega)$. The inset shows the maximal value $P(\omega_m)$ vs $b$.
%}
%  \label{sketch7}
%\end{figure}
For $N = 3$ and a LNT form of $\Psi(\tau)$, we observe essentially the same behavior.

To substantiate our theoretical predictions,
we performed a numerical analysis of $\Psi(\tau)$ and  $P(\omega)$ for  a
quasi 1D
disordered Anderson model defined on a rectangular lattice
of size $L \times W$ (with $L = 100$ and
$W = 3$)
with
two single-channel leads connected to sites $(1,2)$ and $(L,2)$.
Our (isolated) system
is
described  by the  Hamiltonian  $H =  \sum_i
\epsilon_i\ket{i}\bra{i}   +  \sum_{i\ne  j}   t_{ij}  \ket{i}\bra{j}$,
where  $t_{ij}$  are
the  hopping  rates  between the
neighboring sites $i$ and $j$, and $\epsilon_i$ is the energy at the
site $i$, which is a centered, $\delta$-correlated Gaussian
random variable.
%  The
% closed  system  is  described  by  a Hamiltonian  $H$,  defined  on  a
% rectangular lattice of size $L_x\times  L_y$ (and $L_x \gg L_y$) given
% by
% \begin{equation} \label{tight-binding}
% $H = \sum_i \epsilon_i\ket{i}\bra{i} +
% \sum_{i\ne j} t_{ij} \ket{i}\bra{j}$.
% \end{equation}
% We denote by $t_{ij}$ the uniform hopping rate between neighbouring
% sites $i$ and $j$ and by $\epsilon_i$ the disordered on-site energy at
% site $i$, which is a centered Gaussian random variable, independently
% from site to site.
%where $\epsilon_i$ is the on-site energy at the $i$-th site of a
%square lattice of size $L_x\times L_y$ and lattice size $1$, and
%$t_{ij}$ are the hopping strengths between nearest neighbour sites. We
%consider impurity disorder by setting Gaussian distributed random
%energies $\epsilon_i$ and $t_{ij} = 1$ if the distance between sites
%$i$ and $j$ is $1$ and zero otherwise.

\begin{figure}[ht]
  \centerline{\includegraphics*[width=\linewidth]{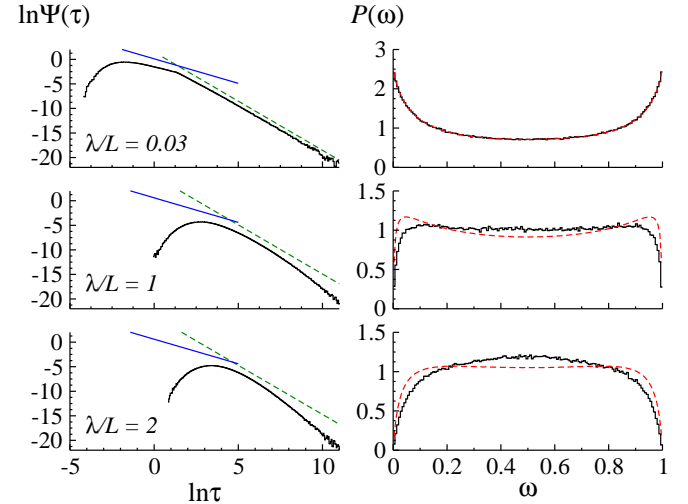}}
  \caption{(Color online) Distributions $\Psi(\tau)$ and $P(\omega)$ for a quasi-1D
    Anderson wire  for different values of $\lambda$: from top to bottom $\lambda = 3$, $100$ and $200$. In
    the left panel we plot $\tau^{-2}$ (dashed green curve) and
    $\tau^{-1}$ (solid blue curve). In the right panel the dashed red
    curve is the corresponding $P(\omega)$ calculated under the assumption of statistical
    independence of the $\tau_k$s.}
  \label{anderson}
\end{figure}

Our numerical results are summarized in Fig.~\ref{anderson}.
In the left panel we depict $\Psi(\tau)$ for different values of the localization length $\lambda \sim 1/\langle \epsilon_i^2
\rangle$.
For
$\lambda/L \ll 1$, one observes that $\Psi(\tau)$ decays asymptotically
as $1/\tau^{2}$, which corresponds to
$\mu=1$. On the other hand, $\Psi(\tau)$ clearly exhibits an intermediate
regime with a slower than $1/\tau^2$ decay $(\mu < 1)$. When
$\lambda$ increases, this intermediate regime shrinks and also
the asymptotic decay becomes faster (possibly, a log-normal).
The right panel shows the corresponding distributions $P(\omega)$, evidencing a
transition from $U$-like
to bell-shaped curves
upon an increase of disorder.
The $U$-like shape ($\lambda/L \ll 1$, top right)
stems out
of the intermediate regime with $\mu < 1$.
Interestingly, the critical distribution $P(\omega) \approx 1$ is observed for $\lambda/L
\simeq 1$ (middle right), \ie, when $\lambda$ is equal to the length of the system.
For $\lambda/L > 1$, a faster than $1/\tau^2$ decay
of $\Psi(\tau)$
leads to a bell-shaped $P(\omega)$.

As a test of statistical
independence of the actual $\tau_k$'s,
we have computed the distribution
$P_{uncor}(\omega)$ (dashed red
curves in Fig.~\ref{anderson}) of the random variable $\tau_1/(\tau_1+\tau_2)$
where
$\tau_1$ and $\tau_2$ are
i.i.d. random
variables drawn from the numerically
observed $\Psi(\tau)$. One notices a good agreement between
$P(\omega)$ (black
histogram) and $P_{uncor}(\omega)$, which is
a clear indication of the lack of correlations between the
different channels for $\lambda/L \ll 1$.
Correlations between
channels induce some discrepancies between
$P(\omega)$ and $P_{uncor}(\omega)$
only for $\lambda/L \gtrsim 1$, when the extension of the typical
eigenfunction becomes of the order of the system size.
Consequently, the scattering exhibits a
transition as the strength of the disorder is varied: $\tau_1$ and
$\tau_2$ are most likely very different for $\lambda/L < 1$ and most
likely the same for $\lambda/L > 1$.  We conjecture that our
findings can be extrapolated to thin 3D disordered
wires, leading to a disproportionate contribution of the open channels
to the total scattering in the diffusive regime, and a proportionate
contribution in the metallic regime.

\begin{figure}[ht]
  \centerline{\includegraphics*[width=\linewidth]{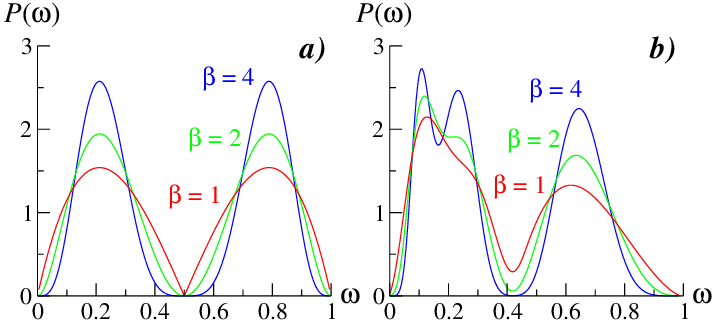}}
  \caption{(Color online) $P(\omega)$ for the proper delay times for different values
    of $\beta$.  {\bf a)}: $N = 2$, Eq.~(\ref{beenaker}). {\bf b)}: $N
    = 3$.  }
  \label{sketch8}
\end{figure}

To summarize, we have studied the distribution $P(\omega)$ of the
random variable $\omega$, Eq.~(\ref{omega}), which defines the
contribution of a given channel to the WDT in a
system with a few open, independent, statistically equivalent
channels.  We have shown that for $2$-channel systems intermediate
power-law tails with $\mu \leq 1$ in the distribution of the partial
delay times entail breaking of the symmetry between the channels;
$P(\omega)$ has a characteristic $U$-shape form and the average
$\langle \omega \rangle = 1/2$ corresponds to the least probable
value.  For $\mu > 1$ the symmetry is statistically preserved and
$\langle \omega \rangle = 1/2$ is also the most probable value.  For $N = 3$ the symmetry between the
channels is always broken which results in unusual bimodal forms of $P(\omega)$.

Finally, we briefly comment on the effect of correlations on
$P(\omega)$.  We mention two known results on the joint distributions
of the partial and of the proper delay times for which we can evaluate
$P(\omega)$ exactly. The joint distribution function of any two
\textit{partial} delay times in a system with $N$ channels and
arbitrary $\beta$ has been calculated in Ref.~\cite{corr}. From
this result, we compute exactly the distribution $P(\omega)$ for two
statistically equivalent (but not independent) channels:
\begin{equation}
\label{cor}
P(\omega) = D \, \omega^{3 \beta/2} \left(1 - \omega\right)^{3 \beta/2} \, ,
\end{equation}
% B = \frac{\Gamma(2 + 3 \beta)}{\Gamma^2(1 + 3 \beta/2)}
with $D = \frac{\Gamma(2 + 3 \beta)}{\Gamma^2(1 + 3 \beta/2)}$, which
is also a beta-distribution, but with an exponent ($= 3 \beta/2$)
larger than the one ($= \beta/2$) in Eq.~(\ref{lim}) corresponding to
two independent channels. For the same $\beta$, the distribution in
Eq.~(\ref{cor}) is narrower than $P(\omega)$ in Eq.~(\ref{lim}) with
$\mu = 1 + \beta/2$ (see Fig.~\ref{sketch}).  Hence, one may argue
that the partial delay times
\textit{attract} each other, which interaction competes with the
symmetry breaking produced by the intermediate power-law tails. Note, as well, that the larger $\beta$ is,
the narrower is the distribution $P(\omega)$.

The joint distribution of $N$ \textit{proper} delay
times in a system with $N$ open channels is also known
exactly~\cite{beenaker}.  It turns out to be given by the Laguerre
ensemble of random-matrix theory and is defined as a product of
$\prod_{k = 1}^N\Psi(\tau'_k)$, where each $\Psi(\tau_k')$ as
in~(\ref{distribution}) with $\mu = 1 + N \beta/2$, times the Dyson's
circular ensemble, $\prod_{i < j} |1/\tau_i' -
1/\tau_j'|^{\beta}$. Due to the latter factor, the $\tau_k'$s harshly
\textit{repel} each other.  For $N = 2$, we obtain
\begin{equation}
\label{beenaker}
P(\omega) = F \, \omega^{\beta} (1 - \omega)^{\beta}  |1 - 2 \omega|^{\beta},
\end{equation}
% V = \frac{4^{1+\beta} \Gamma\left(\frac{5 + 3 \beta}{2}\right)}{3 \Gamma\left(\frac{1 + \beta}{2}\right) \Gamma(2 + \beta)}
where $F$ is a computable normalization constant.  Remarkably,
$P(\omega)$ in~(\ref{beenaker}) is a product of the beta-distribution
in Eq.~(\ref{lim}) and a factor $|1 - 2 \omega|^{\beta}$, which is a
new feature here and stems from the correlations between $\tau_1'$ and
$\tau_2'$.  This factor forbids $\tau_1'$ and $\tau_2'$ to have the
same values and enhances the symmetry breaking [see
Fig.~(\ref{sketch8})]. Note, however, that two peaks in $P(\omega)$ become
narrower the larger $\beta$ is.
Finally, for $N=3$, for which we can also
compute $P(\omega)$ exactly, one shows that a combined effect of the
repulsion and of the intermediate power-law tail results in a very
peculiar asymmetric structure of the distribution [see Fig.~(\ref{sketch8})],
which becomes increasingly more complicated when $\beta$ increases.

%The expression for $P(\omega)$ in the $3$-channel case appears
%too lengthy to be presented here and we only plot
%the exact result in Fig.~(\ref{sketch8}). One realizes that
%for $N = 3$ a combined effect of the repulsion and of the
%intermediate power-law tail results in a very peculiar
%structure of the distribution revealing a particularly
%strong symmetry breaking between the channels.

We thank Y. V. Fyodorov, J. A. M\'endez-Berm\'udez and
T. Kottos for very helpful discussions and comments.

%%%%%%%%%%
%%%%%%%%%%


\begin{thebibliography}{99}


\bibitem{yan1} Y. V. Fyodorov and H.-J. Sommers, J. Math. Phys. {\bf
    38}, 1918 (1997); T.~Kottos, J. Phys. A {\bf 38}, 10761 (2005).

\bibitem{wigner} E. P. Wigner, Phys. Rev. {\bf 98}, 145 (1955);
F. T. Smith, Phys. Rev. {\bf 118}, 349 (1960).

\bibitem{corr} D. V. Savin, Y. V. Fyodorov, and H.-J. Sommers,
  Phys. Rev. E {\bf 63}, R035202 (2001).

\bibitem{beenaker} P. W. Brouwer, K. M. Frahm and C. W. Beenakker,
  Phys. Rev. Lett. {\bf 78}, 4737 (1997).


%\bibitem{comtet1} A. Comtet and C. Texier, J. Phys. A {\bf 30}, 8017 (1997).

\bibitem{comtet2} C. Texier and A. Comtet, J. Phys. A {\bf 30}, 8017
  (1997); Phys. Rev. Lett. {\bf 82}, 4220 (1999).

\bibitem{bolton} C. J. Bolton-Heaton et al.,
%C. J. Lambert, V. I. Falco, V. Prigodin and A. J. Epstein,
  Phys. Rev. B {\bf 60}, 10569 (1999).

\bibitem{ossipov} A. Ossipov, T. Kottos, and T. Geisel, Phys. Rev. B
  {\bf 61}, 11411 (2000).

\bibitem{fJLett} Y.V. Fyodorov, JETP Letters {\bf 78}, 250 (2003).

\bibitem{3} A. Ossipov and Y. V. Fyodorov, Phys. Rev. B {\bf 71},
  125133 (2005).

\bibitem{steinbach} F. Steinbach, A. Ossipov, T. Kottos, and
  T. Geisel, Phys. Rev. Lett. {\bf 85}, 4426 (1999).

%\bibitem{steinbach} F. Steinbach {\it et al.}, Phys. Rev. Lett. {\bf 85}, 4426 (1999).

\bibitem{geisel} A. Ossipov, T. Kottos, and T. Geisel,
  Europhys. Lett. {\bf 62}, 719 (2003).

\bibitem{fyod} Y. V. Fyodorov, D. V. Savin, and H.-J. Sommers,
  Phys. Rev. E {\bf 55}, R4857 (1997).

\bibitem{sym} Y. V. Fyodorov and H.-J. Sommers, Phys. Rev. Lett. {\bf
    76}, 4709 (1996).

\bibitem{2} V. A. Gopar, P. A. Melo, and M. B\"uttiker,
  Phys. Rev. Lett. {\bf 77}, 3005 (1996).

\bibitem{redner} G. Oshanin and S. Redner, Europhys. Lett. {\bf 85},
  10008 (2009); G. Oshanin and G. Schehr, arXiv:1005.1760v1.

\bibitem{we_jstat} C. Mej\'ia-Monasterio, G. Oshanin and G. Schehr,
  J. Stat. Mech. P06022 (2011).

\end{thebibliography}
\end{document}